\documentclass[journal]{IEEEtran} 

\usepackage{cite}

\ifCLASSINFOpdf
  \usepackage[pdftex]{graphicx}

  \DeclareGraphicsExtensions{.pdf,.jpeg,.png}
\else
   \usepackage[dvips]{graphicx}

   \DeclareGraphicsExtensions{.eps}
\fi

\usepackage[cmex10]{amsmath}

\usepackage{array}
\usepackage{mdwmath}
\usepackage{mdwtab}

\usepackage{eqparbox}

\usepackage[caption=false,font=footnotesize]{subfig}

\usepackage{multirow}
\usepackage{amsfonts}
\usepackage{rotating}
\usepackage{balance}
\usepackage{tikz}
\usepackage{pgfplots}
\newcommand{\var}{{\mathbb V}\text{ar}}
\newcommand{\sign}{\text{sign}}

\begin{document}
%
\title{Optimal Asymmetric Binary Quantization for Estimation Under Symmetrically Distributed Noise}
%
%
%
\author{Rodrigo~Cabral Farias*%
             ,~Eric~Moisan~
             and~Jean-Marc~Brossier~
\thanks{The authors are with Grenoble Laboratory of Images, Speech, Signal and Automatics, Department of Images and Signals, 38402 Saint Martin d'H\`eres, France (e-mail: rodrigo.cabral-farias@gipsa-lab.grenoble-inp.fr; jean-marc.brossier@gipsa-lab.grenoble-inp.fr; eric.moisan@gipsa-lab.grenoble-inp.fr). This work was supported by the Erasmus Mundus EBWII program.}}
\markboth{}%
{Shell \MakeLowercase{}}
%



\maketitle

\begin{abstract}
Estimation of a location parameter based on noisy and binary quantized measurements is considered in this letter.  We study the behavior of the Cram\'er--Rao bound as a function of the quantizer threshold for different symmetric unimodal noise distributions. We show that, in some cases, the intuitive choice of threshold position given by the symmetry of the problem, placing the threshold on the true parameter value, can lead to locally worst estimation performance.
\end{abstract}


%
\IEEEpeerreviewmaketitle


\section{Introduction}

Reduction of production costs of sensors and communication devices lead to the emergence of a new signal processing area, sensor networks. In a sensor network, a large number of sensors is used to gather information on some physical quantities (pressure, humidity, temperature) or to detect the occurrence of some events (dam failure, fire) \cite{Puccinelli2005}. Due to the large number of sensors, which in some cases communicate wirelessly, the communication rate available for transmitting the sensor measurements is constrained. A possible way for the sensors to communicate under this constraint is to consider that the measurements are coarsely quantized, for example using a one bit quantizer.

This is the background motivation for this work, in which we study estimation based on binary quantized measurements. In a more detailed way, the estimation problem is the following: estimate the scalar parameter $ x\in\mathbb{R} $, from $ N $ noisy measurements $ \left\lbrace Y_{k}=x+V_{k},\,k=1,\,\cdots,\,N \right\rbrace  $, which are quantized $\left\lbrace i_{k}=Q\left( Y_k \right),\,k=1,\,\cdots,\,N\right\rbrace $. The noise $ \left\lbrace V_{k},\,k=1,\,\cdots,\,N \right\rbrace $ is a sequence of real, independent and identically distributed random variables and $Q$ is the binary quantizer $Q(Y_{k})=\sign(Y_{k}-\tau_{0})\in\left\{ -1,+1 \right\}$.

The specific objective of this work is to characterize how the quantizer threshold $ \tau_{0} $ affects the estimation performance and what its optimal value for different noise distributions is. In what follows, $ F $ denotes the cumulative distribution function (CDF) of the noise and $ f $ its probability density function (PDF). We assume that $ f\left(v\right)  $ is a strictly positive function with even symmetry and decreasing for $ v>0 $. Specific distributions for which these assumptions hold are the Gaussian distribution, the Cauchy distribution, the Laplacian distribution and all the generalized Gaussian class of distributions.

\section{Estimation and asymptotic variance}
\label{est_var}
From the definition of the problem, the binary outputs have the following probabilities
\begin{equation}
i_{k}=\begin{cases}
+1,\quad \mathbb{P}\left(Y_{k}>\tau_{0} \right)= \mathbb{P}\left(x+V_{k}>\tau_{0} \right) =1-p,\\
-1,\quad \mathbb{P}\left(Y_{k}\leq\tau_{0} \right)= \mathbb{P}\left(x+V_{k}\leq\tau_{0} \right)= F\left(\tau_{0}-x \right) ,
\end{cases}
\label{eq_ik_prob}
\end{equation}
\noindent thus, if we knew $ \tau_{0} $, $ F $ and $ p $ we could obtain $ x $ with
\begin{equation}
x=\tau_{0}-F^{-1}\left(p\right).
\label{eq_x_F_inv}
\end{equation}
This indicates that if we have an accurate estimate $ \hat{p} $ of $ p $, we can use it to obtain an estimate $ \hat{X} $ of $ x $
\begin{equation}
\hat{X}=\tau_{0}-F^{-1}\left(\hat{p}\right).
\label{eq_x_F_inv}
\end{equation}
\noindent Now, note that $ \frac{1-i_{k}}{2} $ is a Bernoulli variable with parameter $ p $. As a consequence, an efficient estimator of $ p $, which can be used in (\ref{eq_x_F_inv}), is $\hat{p}=\frac{1}{N}\sum\limits_{k=1}^{N}\frac{1-i_{k}}{2} $. Observe that, for large $ N $,  the error $ \epsilon_{p} $ of this estimator is approximately normal with zero mean and variance $ \frac{p\left(1-p\right) }{N} =\frac{F\left(\tau_{0}-x\right)\left[ 1-F\left(\tau_{0}-x\right) \right]  }{N}$.

To obtain the asymptotic variance of this estimator, we can study how a small error $ \epsilon_{p} $ on the estimation of $ p $ is related to a small error $ \epsilon_{x} $ on the estimation of $ x $. We can use a first order approximation of $p+\epsilon_{p}= F\left(\tau_{0}-x-\epsilon_{x} \right) $ around $ x $ to obtain
\begin{equation}
 p+\epsilon_{p}\approx  F\left(\tau_{0}-x \right)+\epsilon_{x} \frac{\text{d} F\left(\tau_{0}-x\right) }{\text{d}x}=p-\epsilon_{x}f\left(\tau_{0}-x \right).
\label{eq_deviation_1}
\end{equation}
This gives $ \epsilon_{x}\approx -\frac{1}{f\left(\tau_{0}-x\right) }\epsilon_{p} $, indicating that for large $ N $
\begin{equation}
\var\left( \epsilon_{x}\right) \approx\frac{\var\left(\epsilon_{p}\right) }{f^{2}\left(\tau_{0}-x\right) }=\frac{1}{N}\frac{F\left(\tau_{0}-x\right)\left[ 1-F\left(\tau_{0}-x\right) \right]  }{f^{2}\left(\tau_{0}-x\right)}.
\label{eq_var}
\end{equation}
In fact, it can be shown \cite{Papadopoulos2001} that the simple and intuitive estimator of $ x $ defined above is the maximum likelihood estimator (MLE) and its asymptotic variance is equal to the Cram\'er--Rao bound (CRB). This means that the estimator is asymptotically optimal in terms of variance compared with other unbiased estimators. 

The asymptotic variance and the CRB are a function of the difference $\varepsilon=\tau_{0}-x $ between the parameter and threshold values
\begin{equation}
\text{CRB}\left(\varepsilon\right) =N^{-1 } B\left(\varepsilon\right) 
\mbox{ with }
B\left(\varepsilon\right)=\frac{F\left(\varepsilon \right)\left[1-F\left(\varepsilon \right) \right]
}{f^{2}\left(\varepsilon\right) }.
\label{eq_b}
\end{equation}
\noindent Therefore, we can directly study the influence of the threshold position on estimation performance by analyzing $ B\left(\varepsilon\right)  $. This is what we will do in the next sections for different symmetric unimodal noise distributions (distributions with $ F\left(v\right)  $ convex for $ v<0 $ and $ F\left(v\right)  $ concave for $ v>0 $).

\section{Commonly used symmetric noise distributions}

\paragraph{Gaussian}a usual case of unimodal distribution is the Gaussian distribution $ f_{G}\left(v\right)=\left[\delta\sqrt{\pi}\right]^{-1}\exp\left[-v^{2}/\delta^{2}\right] $, where $ \delta $ is a scale parameter. This case is analyzed in \cite{Papadopoulos2001} and \cite{Ribeiro2006a}, where it is shown that the minimum value of $ B\left(\varepsilon\right)  $ is obtained for $ \varepsilon=0$ and that $ B\left(\varepsilon\right) $ increases exponentially to infinity when $ \left\vert \varepsilon \right\vert $ increases to infinity. Also, it is shown that the minimum relative loss of estimation performance with respect to (w.r.t.) the MLE with continuous measurements is $ \frac{\pi}{2} $ or approximately $ 2 \text{dB} $ (decibel), which is a small loss if we consider how coarse the information that we have is. 

\paragraph{Cauchy}a standard family of distributions, that is used as noise model in robust statistics, is the Student's-t family \cite{Lange1989}, a specific member of this family that is used to model impulsive noise is the heavy-tailed Cauchy distribution $ f_{C}\left(v\right)=\frac{1}{\pi\delta}\frac{1}{\left[ 1+\left(\frac{v}{\delta}\right)^{2}\right] } $. For this distribution, if we evaluate $ B\left(\varepsilon\right)  $, we find a behavior similar to the Gaussian case: $ \varepsilon=0 $ is the minimum point and $ B\left( \varepsilon\right)  $ increases with $ \left\vert \varepsilon \right\vert $. In this case, the relative loss w.r.t. the continuous measurement estimator is $ \frac{\pi^{2}}{8} $ or approximately $ 0.9 \text{dB}$, which is even smaller than the Gaussian loss.

\paragraph{Laplacian}another common distribution that is used to model impulsive noises is the Laplacian distribution $ f_{L}\left(v\right)=\frac{1}{2\delta}\exp\left( -\left\vert \frac{v}{\delta} \right\vert\right)  $. For this distribution $ B\left(\varepsilon\right)  $ in (\ref{eq_b}) can be calculated analytically
\begin{equation}
B\left(\varepsilon\right) =\delta^{2}\left[2\exp\left( \left\vert \frac{\varepsilon}{\delta} \right\vert \right) -1 \right]
\label{eq_b_laplace}
\end{equation}
\noindent and we can easily see the same type of pattern as $ \varepsilon=0 $ is the minimum and the function increases exponentially with $ \left\vert \frac{\varepsilon}{\delta} \right\vert $. Note that the optimal performance now is $ CRB\left(0\right)=\frac{\delta^2}{N}  $, which is exactly equal to the CRB for continuous measurements, therefore well-tuned binary quantization, in this case, causes no loss of performance. 

In all cases above, the optimal threshold is $ \tau_{0}=x $, optimal quantization is symmetric, \textit{i.e.} with equiprobable outputs. As $ x $ is unknown, a possible way to achieve in practice the small performance losses indicated above is to use its best approximation based on the measurements which is $ \hat{X} $, this gives rise to the adaptive approaches proposed in \cite{Papadopoulos2001}, \cite{Fang2008} and \cite{Farias2013}.

From the examples above, intuition seems to indicate that the symmetric behavior of optimal quantization may be a characteristic that can be generalized to all symmetric unimodal distributions \cite{Wang2010}. Unfortunately, our intuition is wrong in this case, not only this is not true for all unimodal distributions, but it happens that $ \tau_{0}=x $ can be locally the worst choice for the quantizer threshold. 

\section{Local condition for symmetry}
To prove that our intuition is wrong, we study the local behavior of $ B\left(\varepsilon\right)  $ around $ \varepsilon=0 $. We denote  $f^{\left(n\right) }$ the $ n $-th order derivative of $ f $ and we impose $ f^{\left(1\right) }\left(0\right)=0$.

For symmetric unimodal distributions $f\left(\varepsilon\right)=f\left(-\varepsilon\right)$ and
$F\left(\varepsilon\right)=1-F\left(-\varepsilon\right)$, thus $ B\left(\varepsilon\right)$ has even symmetry. 

The first derivative of $ B\left(\varepsilon\right)  $ w.r.t. $ \varepsilon $ is
\begin{equation}
\frac{\text{d}B}{\text{d}\varepsilon}=\frac{f^{2}\left(\varepsilon\right)\left[1-2F\left(\varepsilon\right) \right]-2F\left(\varepsilon\right)\left[1-F\left(\varepsilon\right)  \right]f^{\left(1\right) }\left(\varepsilon\right)   }{f^{3}\left(\varepsilon\right) } ,\nonumber
\end{equation}
at $ \varepsilon=0 $ we have $1-2F\left(0\right)=0$, and only the second term is non-zero. As $ f^{\left(1\right) }\left(0\right)=0$, $ \frac{\text{d}B}{\text{d}\varepsilon}=0 $ at $ \varepsilon=0 $ and this is an extremum point. To verify if it is a minimum or a maximum, we calculate the second derivative:
\begin{eqnarray}
  \frac{\text{d}^{2}B}{\text{d}\varepsilon^{2}}=-2+\frac{1}{f^{4}\left(\varepsilon\right)}\times\left\lbrace -3f^{2}\left(\varepsilon\right)f^{\left(1\right) }\left(\varepsilon\right)\left[1-2F\left(\varepsilon\right) \right]+ \right.\nonumber\\
  \qquad\qquad\left. +F\left(\varepsilon\right)\left[1-F\left(\varepsilon\right)  \right]\left[6f^{\left(1\right)\,2 }\left(\varepsilon\right)-2f\left(\varepsilon\right)f^{\left(2\right) }\left(\varepsilon\right) \right]  \right\rbrace.\nonumber
\end{eqnarray} 
Therefore, at $ \varepsilon=0 $ we have $\left.\frac{\text{d}^{2}B}{\text{d}\varepsilon^{2}}\right\vert_{\varepsilon=0}=-\frac{1}{2}\frac{f^{\left(2\right)}\left(0\right)
}{f^{3}\left(0\right) }-2$ and a condition to have a local minimum is
\begin{equation}
-f^{\left(2\right) }\left(0\right)>4f^{3}\left(0\right).
\label{cond_sec_dev}
\end{equation}
This condition can be easily verified in the Gaussian and Cauchy cases, however, as we present next, it is not true for all symmetric unimodal distributions. 

\begin{figure}[t!]
\centering
\begin{tikzpicture}

\begin{axis}[
hide x axis, hide y axis,
width=2.72in,
height=1.7in,
scale only axis,
xmin=-5, xmax=5,
xlabel={$\varepsilon$},
ymin=-0.05, ymax=0.4,
axis on top]
\addplot [
color=black,
solid
]
coordinates{
 (-4.5,9.56652948729823e-05)(-4.45,0.000116699888947094)(-4.4,0.000142004051329242)(-4.35,0.000172363494623096)(-4.3,0.000208691187536118)(-4.25,0.000252044481968382)(-4.2,0.000303643886171529)(-4.15,0.000364893520980511)(-4.1,0.000437403271518954)(-4.05,0.000523012616803859)(-4,0.000623816083959761)(-3.95,0.000742190232011196)(-3.9,0.000880822022277738)(-3.85,0.00104273837820396)(-3.8,0.00123133667713537)(-3.75,0.00145041585040773)(-3.7,0.00170420769667546)(-3.65,0.00199740793742943)(-3.6,0.00233520646417413)(-3.55,0.00272331614505824)(-3.5,0.00316799947648038)(-3.45,0.00367609228422079)(-3.4,0.00425502360116512)(-3.35,0.00491283077715101)(-3.3,0.00565816881359007)(-3.25,0.0065003128642174)(-3.2,0.00744915280667066)(-3.15,0.0085151787707688)(-3.1,0.00970945651152093)(-3.05,0.0110435915411477)(-3,0.0125296809876521)(-2.95,0.0141802522303514)(-2.9,0.0160081874774824)(-2.85,0.0180266335991775)(-2.8,0.0202488967117876)(-2.75,0.0226883212269198)(-2.7,0.0253581533299943)(-2.65,0.0282713891369565)(-2.6,0.031440608091283)(-2.55,0.0348777925027626)(-2.5,0.038594134489734)(-2.45,0.0425998319614124)(-2.4,0.0469038756594359)(-2.35,0.0515138296596044)(-2.3,0.0564356081068899)(-2.25,0.0616732513093777)(-2.2,0.0672287046395503)(-2.15,0.0731016039737055)(-2.1,0.0792890716317662)(-2.05,0.0857855269500815)(-2,0.0925825157194663)(-1.95,0.0996685627410947)(-1.9,0.107029051686664)(-1.85,0.114646136290839)(-1.8,0.122498686649625)(-1.75,0.13056227404651)(-1.7,0.138809197279741)(-1.65,0.147208552922484)(-1.6,0.155726351318824)(-1.55,0.164325679411353)(-1.5,0.172966910721798)(-1.45,0.18160796197845)(-1.4,0.190204595019031)(-1.35,0.198710761712933)(-1.3,0.207078988761677)(-1.25,0.215260798371461)(-1.2,0.223207159967428)(-1.15,0.230868967356706)(-1.1,0.238197535066407)(-1.05,0.245145107002538)(-1,0.251665370113249)(-0.95,0.257713965409776)(-0.9,0.263248988512697)(-0.85,0.268231471858339)(-0.8,0.272625840825315)(-0.75,0.276400336325451)(-0.7,0.279527396843882)(-0.65,0.281983993503071)(-0.6,0.283751912454247)(-0.55,0.284817979752838)(-0.5,0.285174224834319)(-0.45,0.285174224834319)(-0.4,0.285174224834319)(-0.35,0.285174224834319)(-0.3,0.285174224834319)(-0.25,0.285174224834319)(-0.2,0.285174224834319)(-0.15,0.285174224834319)(-0.1,0.285174224834319)(-0.05,0.285174224834319)(0,0.285174224834319)(0.05,0.285174224834319)(0.1,0.285174224834319)(0.15,0.285174224834319)(0.2,0.285174224834319)(0.25,0.285174224834319)(0.3,0.285174224834319)(0.35,0.285174224834319)(0.4,0.285174224834319)(0.45,0.285174224834319)(0.5,0.285174224834319)(0.55,0.284817979752838)(0.6,0.283751912454247)(0.65,0.281983993503071)(0.7,0.279527396843882)(0.75,0.276400336325451)(0.8,0.272625840825315)(0.85,0.268231471858339)(0.9,0.263248988512697)(0.95,0.257713965409776)(1,0.251665370113249)(1.05,0.245145107002538)(1.1,0.238197535066407)(1.15,0.230868967356706)(1.2,0.223207159967428)(1.25,0.215260798371461)(1.3,0.207078988761677)(1.35,0.198710761712933)(1.4,0.190204595019031)(1.45,0.18160796197845)(1.5,0.172966910721798)(1.55,0.164325679411353)(1.6,0.155726351318824)(1.65,0.147208552922484)(1.7,0.138809197279741)(1.75,0.13056227404651)(1.8,0.122498686649625)(1.85,0.114646136290839)(1.9,0.107029051686664)(1.95,0.0996685627410947)(2,0.0925825157194663)(2.05,0.0857855269500815)(2.1,0.0792890716317662)(2.15,0.0731016039737055)(2.2,0.0672287046395503)(2.25,0.0616732513093777)(2.3,0.0564356081068899)(2.35,0.0515138296596044)(2.4,0.0469038756594359)(2.45,0.0425998319614124)(2.5,0.038594134489734)(2.55,0.0348777925027626)(2.6,0.031440608091283)(2.65,0.0282713891369565)(2.7,0.0253581533299943)(2.75,0.0226883212269198)(2.8,0.0202488967117876)(2.85,0.0180266335991775)(2.9,0.0160081874774824)(2.95,0.0141802522303514)(3,0.0125296809876521)(3.05,0.0110435915411477)(3.1,0.00970945651152093)(3.15,0.0085151787707688)(3.2,0.00744915280667066)(3.25,0.0065003128642174)(3.3,0.00565816881359007)(3.35,0.00491283077715101)(3.4,0.00425502360116512)(3.45,0.00367609228422079)(3.5,0.00316799947648038)(3.55,0.00272331614505824)(3.6,0.00233520646417413)(3.65,0.00199740793742943)(3.7,0.00170420769667546)(3.75,0.00145041585040773)(3.8,0.00123133667713537)(3.85,0.00104273837820396)(3.9,0.000880822022277738)(3.95,0.000742190232011196)(4,0.000623816083959761)(4.05,0.000523012616803859)(4.1,0.000437403271518954)(4.15,0.000364893520980511)(4.2,0.000303643886171529)(4.25,0.000252044481968382)(4.3,0.000208691187536118)(4.35,0.000172363494623096)(4.4,0.000142004051329242)(4.45,0.000116699888947094)(4.5,9.56652948729823e-05) 
};

\draw[->,>=latex] (axis cs:-4.5,0)-- (axis cs:4.5,0);
\node[below] at (axis cs:4.5,0) {\small $ \varepsilon $};
\draw (axis cs:0.5,-0.01)-- (axis cs:0.5,0.01);
\node[below] at (axis cs:0.5,0) {\small $ \frac{\alpha}{2} $};
\draw (axis cs:-0.5,-0.01)-- (axis cs:-0.5,0.01);
\node[below] at (axis cs:-0.5,0) {\small $- \frac{\alpha}{2} $};
\draw[dashed] (axis cs:-0.5,0)-- (axis cs:-0.5,0.2852);
\draw[dashed] (axis cs:0.5,0)-- (axis cs:0.5,0.2852);
\draw[->,>=latex] (axis cs:-2.5,0.2)-- (axis cs:-1.65,0.1472);
\node[above] at (axis cs:-2.5,0.2) {\small $f_{GL}\left(\varepsilon\right) $};
\draw[->,>=latex] (axis cs:2.5,0.2)-- (axis cs:1.65,0.1472);
\node[above] at (axis cs: 2.5,0.2) {\small $f_{GR}\left(\varepsilon\right) $};
\draw[->,>=latex] (axis cs:0,0.35)-- (axis cs:0,0.2852);
\node[above] at (axis cs: 0,0.35) {\small $f_{U}\left(\varepsilon\right) $};
\draw (axis cs:-1.5,0.1730)-- (axis cs:-0.5,0.1730);
\draw (axis cs:-1.5,0.1670)-- (axis cs:-1.5,0.1790);
\draw (axis cs:-0.5,0.1670)-- (axis cs:-0.5,0.1790);
\node[above] at (axis cs: -1,0.1730) {\small $\sigma $};
\draw (axis cs:1.5,0.1730)-- (axis cs:0.5,0.1730);
\draw (axis cs:1.5,0.1670)-- (axis cs:1.5,0.1790);
\draw (axis cs:0.5,0.1670)-- (axis cs:0.5,0.1790);
\node[above] at (axis cs: 1,0.1730) {\small $\sigma $};
\end{axis}
\end{tikzpicture}
\caption{PDF for the uniform/Gaussian distribution. The center region is uniform with width $ \alpha $, while the left and right sides are Gaussian with standard deviation parameter $ \sigma $.}
\label{fig_unigauss}
\end{figure}
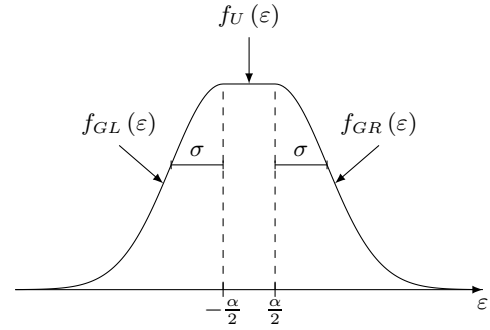

\section{Asymmetric binary quantization}

\subsection{The uniform/Gaussian case}
Condition (\ref{cond_sec_dev}) is false if $f^{\left(2\right) }\left(0\right)=0$, \textit{i.e.} if the noise PDF is flat at the origin. Thus, the first case that comes into mind is the uniform distribution. We can add two Gaussian tails\footnote{The tails need not to be Gaussian, any decreasing tail can be used.} around the uniform distribution to respect the regularity conditions required for the validity of the CRB \cite[p. 67]{Kay1993}. Thus, condition (\ref{cond_sec_dev}) is false for the following PDF (see Fig. \ref{fig_unigauss}):
\begin{equation}
f_{UG}\left(v\right)=
\begin{cases}
f_{GL}\left(v\right)=\frac{1}{C\sqrt{2\pi}\sigma}\exp\left[-\frac{1}{2}\left(\frac{v+\frac{\alpha}{2}}{\sigma}\right)^{2} \right] \\ 
\hspace{3.5cm} \text{pour }v<-\frac{\alpha}{2},\\
f_{U}\left(v\right)=\frac{1}{C\sqrt{2\pi}\sigma}\\
\hspace{3cm} \text{pour }-\frac{\alpha}{2}\leq v\leq \frac{\alpha}{2},\\
f_{GR}\left(v\right)=\frac{1}{C\sqrt{2\pi}\sigma}\exp\left[-\frac{1}{2}\left(\frac{v-\frac{\alpha}{2}}{\sigma}\right)^{2} \right]\\
\hspace{3.5cm} \text{pour } v>\frac{\alpha}{2},
\end{cases} 
\label{unigauss_pdf}
\end{equation} 
\noindent where $ C=1+\frac{\alpha}{\sqrt{2\pi}\sigma} $ is a normalization constant.

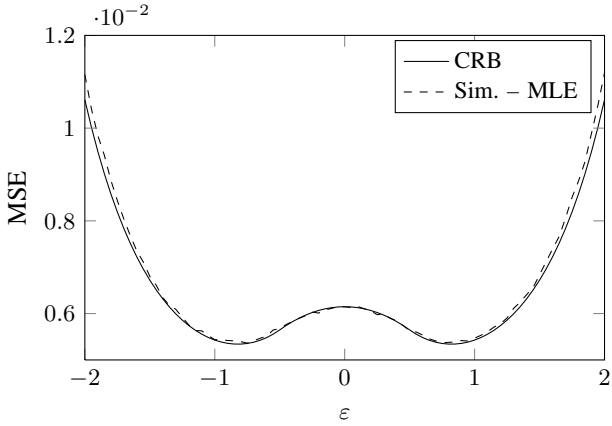
\begin{figure}[t!]
\centering
\begin{tikzpicture}
\begin{axis}[%
width=2.72in,
height=1.7in,
scale only axis,
xmin=-2, xmax=2,
xminorticks=true,
tick label style={font=\small},
label style={font=\small},
xlabel={$\varepsilon$},
ylabel={$\text{MSE}$},
ymin=0.005, ymax=0.012,
legend style={nodes=right,font = \small,at={(axis cs:1.9, 0.0119)}},
ylabel near ticks,
axis on top]
\addplot [
color=black,
solid
]
coordinates{
 (-2,0.010610698917047)(-1.95,0.0100259477721524)(-1.9,0.00949690214690994)(-1.85,0.00901798047802908)(-1.8,0.00858426960184895)(-1.75,0.0081914413162044)(-1.7,0.00783568051659916)(-1.65,0.00751362325355649)(-1.6,0.00722230331345071)(-1.55,0.00695910614037402)(-1.5,0.00672172909842626)(-1.45,0.00650814722785019)(-1.4,0.00631658377935203)(-1.35,0.00614548492265948)(-1.3,0.00599349812116033)(-1.25,0.0058594537471087)(-1.2,0.00574234958373826)(-1.15,0.00564133792370766)(-1.1,0.00555571502938005)(-1.05,0.00548491277105987)(-1,0.0054284923058764)(-0.95,0.0053861397038104)(-0.9,0.00535766346964343)(-0.85,0.00534299395158822)(-0.8,0.00534218467028004)(-0.75,0.00535541564699771)(-0.7,0.0053829988588887)(-0.65,0.00542538600323669)(-0.6,0.0054831788143308)(-0.55,0.00555714224751629)(-0.5,0.00564822092822079)(-0.45,0.00574322092822079)(-0.4,0.00582822092822079)(-0.35,0.00590322092822079)(-0.3,0.00596822092822079)(-0.25,0.00602322092822079)(-0.2,0.00606822092822079)(-0.15,0.00610322092822079)(-0.0999999999999996,0.00612822092822079)(-0.0499999999999998,0.00614322092822079)(0,0.00614822092822079)(0.0499999999999998,0.00614322092822079)(0.0999999999999996,0.00612822092822079)(0.15,0.00610322092822079)(0.2,0.00606822092822079)(0.25,0.00602322092822079)(0.3,0.00596822092822079)(0.35,0.00590322092822079)(0.4,0.00582822092822079)(0.45,0.00574322092822079)(0.5,0.00564822092822079)(0.55,0.00555714224751629)(0.6,0.0054831788143308)(0.65,0.00542538600323669)(0.7,0.0053829988588887)(0.75,0.00535541564699771)(0.8,0.00534218467028004)(0.85,0.00534299395158822)(0.9,0.00535766346964343)(0.95,0.0053861397038104)(1,0.0054284923058764)(1.05,0.00548491277105988)(1.1,0.00555571502938005)(1.15,0.00564133792370766)(1.2,0.00574234958373826)(1.25,0.0058594537471087)(1.3,0.00599349812116033)(1.35,0.00614548492265947)(1.4,0.00631658377935203)(1.45,0.00650814722785019)(1.5,0.00672172909842627)(1.55,0.00695910614037402)(1.6,0.00722230331345071)(1.65,0.0075136232535565)(1.7,0.00783568051659916)(1.75,0.00819144131620441)(1.8,0.00858426960184895)(1.85,0.00901798047802908)(1.9,0.00949690214690994)(1.95,0.0100259477721524)(2,0.010610698917047) 
};

\addlegendentry{$ \text{CRB}$};
\addplot [
color=black,
dashed
]
coordinates{
 (-2,0.0111562967336588)(-1.95,0.0104228475942751)(-1.9,0.00984096204815403)(-1.85,0.00941459472572256)(-1.8,0.00888028885278295)(-1.75,0.00843807507225937)(-1.7,0.00804111006291793)(-1.65,0.00767950220160317)(-1.6,0.00735081816984489)(-1.55,0.00711242977046365)(-1.5,0.00680298754370093)(-1.45,0.00660944823028196)(-1.4,0.00634328331196992)(-1.35,0.00617961616546335)(-1.3,0.00609245435494561)(-1.25,0.00592133126914582)(-1.2,0.00575603863337117)(-1.15,0.00563542176408914)(-1.1,0.00562728089845317)(-1.05,0.00552627491294994)(-1,0.00543786774112812)(-0.95,0.00543625061652057)(-0.9,0.00541945406649729)(-0.85,0.00541107577481573)(-0.8,0.00539408645154372)(-0.75,0.00536355724764158)(-0.7,0.00541272396278853)(-0.65,0.00546926532729465)(-0.6,0.00551218974900395)(-0.55,0.00564638476189321)(-0.5,0.00567661823058195)(-0.45,0.00576359173811614)(-0.4,0.00583428526295682)(-0.35,0.00589880324519885)(-0.3,0.00595594340830412)(-0.25,0.0060314611555686)(-0.2,0.00601165471629408)(-0.15,0.00607931601906748)(-0.0999999999999996,0.00611237412832196)(-0.0499999999999998,0.00615311691354481)(0,0.00614543455449596)(0.0499999999999998,0.00613209620170577)(0.0999999999999996,0.00615387739394611)(0.15,0.00610481890080557)(0.2,0.00605919031427922)(0.25,0.00597737697877911)(0.3,0.00598611352838297)(0.35,0.00589963161600576)(0.4,0.00584532992362206)(0.45,0.0057409563142125)(0.5,0.00563096241691483)(0.55,0.00560275479603267)(0.6,0.00553646559166822)(0.65,0.00545744520112216)(0.7,0.00541062716287432)(0.75,0.00537190764160182)(0.8,0.00537928230052128)(0.85,0.00538952691359357)(0.9,0.00541810851155085)(0.95,0.00540790487719719)(1,0.00546975749408972)(1.05,0.0055335742285489)(1.1,0.0056093064149683)(1.15,0.0057106188673229)(1.2,0.00578414588314304)(1.25,0.00590469679770618)(1.3,0.00611259796562388)(1.35,0.00624351797670885)(1.4,0.00640622373519128)(1.45,0.00656007733893946)(1.5,0.00683718619929842)(1.55,0.0071169796399467)(1.6,0.00739250739173087)(1.65,0.00765731287782422)(1.7,0.00806010692001907)(1.75,0.00850646575203551)(1.8,0.00886118551302855)(1.85,0.00927992495929155)(1.9,0.00989160763933902)(1.95,0.0105053222000118)(2,0.0111952680971741) 
};
\addlegendentry{$ \text{Sim. -- MLE} $};
\end{axis}
\end{tikzpicture}
\caption{$ \text{CRB}$ and simulated MLE MSE for uniform/Gaussian noise. Both the bound and simulated MSE were evaluated for a number of samples $ N=500 $ and for $ \varepsilon $ in the interval $ \left[-2,\;2 \right] $. The MSE for the MLE was evaluated using $ 10^{5} $ realizations of the sample blocks. We considered the following noise parameters: $ \alpha=1 $ and $ \sigma=1 $. }
\label{fig_unigauss_bound}
\end{figure}

Fig. \ref{fig_unigauss_bound} shows $B\left( \varepsilon\right) $ for this PDF. To verify that the bound represents correctly the behavior of the estimation variance, we simulated the MLE $ 10^{5} $ times for blocks of size $ N=500 $ with $ \alpha=1 $ and $ \sigma=1 $. The simulation results were used to evaluate the mean squared error (MSE). As expected, the bound and the simulated MSE are concave around $ \varepsilon=0$, therefore, setting the threshold to the parameter value is locally the worst choice. There are two symmetric minima around $ \varepsilon=0$, as a consequence, optimal quantization is asymmetric, with the threshold being shifted to the right or the left of the true parameter value slightly more than $ \frac{\alpha}{2} $.

\subsection{The generalized Gaussian case}

We can also look for less straightforward cases, without the central uniform behavior, for which (\ref{cond_sec_dev}) is false, for example, the family of generalized Gaussian distributions (GGD) \cite{Varanasi1989}\footnote{$\Gamma\left(x\right)=\int\limits_{0}^{+\infty} z^{x-1} \exp\left(-z \right)\,\text{d}z$ is the Gamma function and \\$\gamma\left(x,w\right)=\int\limits_{0}^{w} z^{x-1} \exp\left(-z \right)\,\text{d}z$ is the incomplete Gamma function.} $f_{GGD}\left(v\right)=\frac{\beta}{2\delta\Gamma\left(\frac{1}{\beta}\right)}\exp\left( -\left\vert \frac{v}{\delta} \right\vert^{\beta}\right)$. $ \beta>0 $ is a shape parameter that allows to control the flatness of the distribution around zero.

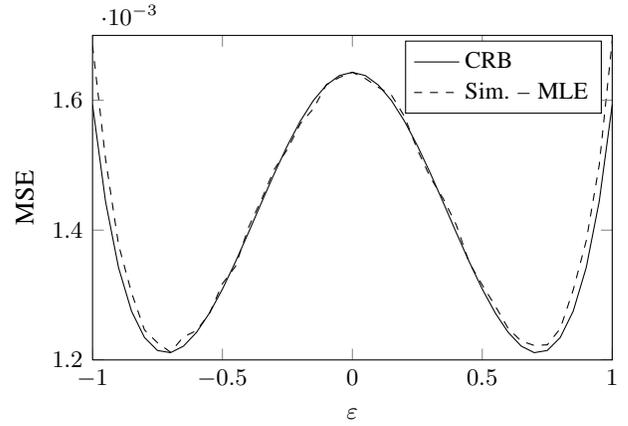
\begin{figure}[t!]
\centering
\begin{tikzpicture}
\begin{axis}[
width=2.72in,
height=1.7in,
scale only axis,
xmin=-1, xmax=1,
xminorticks=true,
tick label style={font=\small},
label style={font=\small},
xlabel={$\varepsilon$},
ylabel={$\text{MSE}$},
ylabel near ticks,
xlabel={$\varepsilon$},
ymin=0.0012, ymax=0.0017,
legend style={nodes=right,font = \small,at={(axis cs:0.96, 0.001692)}},
axis on top]
\addplot [
color=black,
solid
]
coordinates{
 (-1,0.00159327555450052)(-0.95,0.00144432976216789)(-0.9,0.00134242230441552)(-0.85,0.00127526002646941)(-0.8,0.0012345635138319)(-0.75,0.00121458821974427)(-0.7,0.00121120226523196)(-0.65,0.00122129295542604)(-0.6,0.00124237350038099)(-0.55,0.00127231664580045)(-0.5,0.0013091734403446)(-0.45,0.00135105369620348)(-0.4,0.00139605529264292)(-0.35,0.00144223520278599)(-0.3,0.00148761757011191)(-0.25,0.00153023430328371)(-0.2,0.00156819226455248)(-0.15,0.001599758947792)(-0.0999999999999999,0.00162345635959935)(-0.0499999999999998,0.00163815139008898)(0,0.00164313090082461)(0.0499999999999998,0.00163815139008898)(0.0999999999999999,0.00162345635959935)(0.15,0.001599758947792)(0.2,0.00156819226455248)(0.25,0.00153023430328371)(0.3,0.00148761757011191)(0.35,0.00144223520278599)(0.4,0.00139605529264292)(0.45,0.00135105369620348)(0.5,0.0013091734403446)(0.55,0.00127231664580045)(0.6,0.00124237350038099)(0.65,0.00122129295542604)(0.7,0.00121120226523196)(0.75,0.00121458821974427)(0.8,0.0012345635138319)(0.85,0.00127526002646941)(0.9,0.00134242230441552)(0.95,0.00144432976216789)(1,0.00159327555450052) 
};
\addlegendentry{$ \text{CRB}$};
\addplot [
color=black,
dashed
]
coordinates{
 (-1,0.00168476652833973)(-0.95,0.0015098228441049)(-0.9,0.00137786493457174)(-0.85,0.00130371811121787)(-0.8,0.0012461836459264)(-0.75,0.00122646271691635)(-0.7,0.00121198697209537)(-0.65,0.00123556177688848)(-0.6,0.00124513941460583)(-0.55,0.00127162799316274)(-0.5,0.00131747859314082)(-0.45,0.00134415776573316)(-0.4,0.00140393893935028)(-0.35,0.00144695523433206)(-0.3,0.00149395622949402)(-0.25,0.00152456584210045)(-0.2,0.00156417224618717)(-0.15,0.0015870019006387)(-0.0999999999999999,0.00162416796598567)(-0.0499999999999998,0.00163526653171858)(0,0.00164319496348708)(0.0499999999999998,0.00163351862142378)(0.0999999999999999,0.00162005709083707)(0.15,0.00160823901518605)(0.2,0.00157551723735216)(0.25,0.00152413335919248)(0.3,0.00148120451674285)(0.35,0.00144820391216983)(0.4,0.00140794105033534)(0.45,0.00134955937744577)(0.5,0.00131540537301891)(0.55,0.00128318121593163)(0.6,0.00124830399889132)(0.65,0.00122902732922159)(0.7,0.00122233134778781)(0.75,0.00122325251362117)(0.8,0.00124826253956496)(0.85,0.00130808042884512)(0.9,0.00138368918109303)(0.95,0.00150037274160141)(1,0.00169188867613807) 
};
\addlegendentry{$ \text{Sim. -- MLE} $};
\end{axis}
\end{tikzpicture}
\caption{$ \text{CRB} $ and simulated MLE MSE for GGD noise. Both the bound and simulated MSE were evaluated for a number of samples $ N=500 $ and for $ \varepsilon $ in the interval $ \left[-1,\;1 \right] $. The MSE for the MLE was evaluated using $ 10^{5} $ realizations of the sample blocks. We considered the following noise parameters: $ \beta=4 $ and $ \delta=1 $. }
\label{fig_ggd_bound}
\end{figure}

The first derivative of $ f_{GGD} $ at zero is not defined for $\beta\leq 1$, thus we are constrained to analyze the cases where $ \beta>1 $. The second derivative of $ f_{GGD}  $ at zero is $ -\infty $ for $1<\beta<2$. For the Gaussian case ($\beta=2$) the second derivative is negative and respects (\ref{cond_sec_dev}). However, for $\beta>2$, the second derivative is zero and $\varepsilon=0$ is a maximum point. $ B\left(\varepsilon\right)  $ for the GGD is given by
\begin{equation}
B\left(\varepsilon\right)=\frac{\delta^{2}\Gamma^{2}\left(\frac{1}{\beta}\right) }{\beta^{2}}\left[1-\frac{\gamma^{2}\left(\frac{1}{\beta},\left\vert \frac{\varepsilon}{\delta} \right\vert^{\beta} \right) }{\Gamma^{2}\left(\frac{1}{\beta} \right) }  \right] \exp\left(2\left\vert \frac{\varepsilon}{\delta} \right\vert^{\beta} \right).
\label{ggd_bound}
\end{equation}
The shape of this function is shown in Fig. \ref{fig_ggd_bound} through the CRB for $ \beta=4 $, $ \delta=1 $ and $ N=500 $. The simulated performance of $ 10^{5} $ realizations of the MLE are also shown in this figure.
The bound is also close to the estimation performance in this case. Here again, the optimal threshold position must be shifted away from the parameter and the optimal shift depends on the noise scale factor $ \delta $. This shows that the optimal quantizer depends not only on information about the parameter, but also on information about the noise distribution.

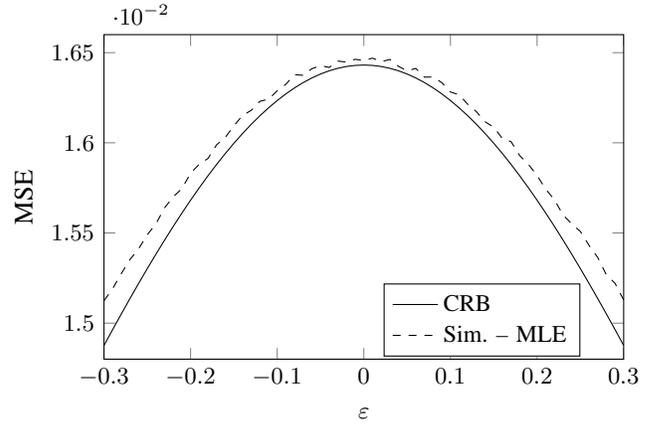
\begin{figure}[t!]
\centering
\begin{tikzpicture}
\begin{axis}[%
width=2.72in,
height=1.7in,
scale only axis,
xminorticks=true,
tick label style={font=\small},
label style={font=\small},
xlabel={$\varepsilon$},
ylabel={$\text{MSE}$},
ylabel near ticks,
xmin=-0.3, xmax=0.3,
ymin=0.0148, ymax=0.0166,
legend style={nodes=right,font = \small,at={(axis cs:0.25, 0.01522)}},
axis on top]
\addplot [
color=black,
solid
]
coordinates{
 (-0.3,0.0148761757011191)(-0.29,0.0149642432802944)(-0.28,0.0150510469048433)(-0.27,0.0151364326037525)(-0.26,0.0152202480882865)(-0.25,0.0153023430328371)(-0.24,0.0153825693641055)(-0.23,0.0154607815572183)(-0.22,0.0155368369373353)(-0.21,0.0156105959852586)(-0.2,0.0156819226455248)(-0.19,0.0157506846354258)(-0.18,0.0158167537533899)(-0.17,0.0158800061851348)(-0.16,0.015940322806007)(-0.15,0.01599758947792)(-0.14,0.0160516973393223)(-0.13,0.0161025430866464)(-0.12,0.0161500292457243)(-0.11,0.0161940644316966)(-0.1,0.0162345635959935)(-0.09,0.0162714482590289)(-0.08,0.0163046467273177)(-0.07,0.0163340942938053)(-0.06,0.016359733420287)(-0.05,0.0163815139008898)(-0.04,0.0163993930056913)(-0.03,0.0164133356036602)(-0.02,0.0164233142642178)(-0.01,0.0164293093368403)(0,0.0164313090082461)(0.01,0.0164293093368403)(0.02,0.0164233142642178)(0.03,0.0164133356036602)(0.04,0.0163993930056913)(0.05,0.0163815139008898)(0.06,0.016359733420287)(0.07,0.0163340942938053)(0.08,0.0163046467273177)(0.09,0.0162714482590289)(0.1,0.0162345635959935)(0.11,0.0161940644316966)(0.12,0.0161500292457243)(0.13,0.0161025430866464)(0.14,0.0160516973393223)(0.15,0.01599758947792)(0.16,0.015940322806007)(0.17,0.0158800061851348)(0.18,0.0158167537533899)(0.19,0.0157506846354258)(0.2,0.0156819226455248)(0.21,0.0156105959852586)(0.22,0.0155368369373353)(0.23,0.0154607815572183)(0.24,0.0153825693641055)(0.25,0.0153023430328371)(0.26,0.0152202480882865)(0.27,0.0151364326037525)(0.28,0.0150510469048433)(0.29,0.0149642432802944)(0.3,0.0148761757011191) 
};
\addlegendentry{$ \text{CRB} $};
\addplot [
color=black,
dashed
]
coordinates{
 (-0.3,0.0151240715648103)(-0.29,0.0151911322526223)(-0.28,0.0152747090202747)(-0.27,0.0153526388581029)(-0.26,0.0154122679494791)(-0.25,0.0154903618969984)(-0.24,0.0155507928644868)(-0.23,0.0156397075930312)(-0.22,0.0157007190957411)(-0.21,0.0157458948858329)(-0.2,0.0158321816925334)(-0.19,0.0158812609557514)(-0.18,0.0159128671818378)(-0.17,0.0159889291040085)(-0.16,0.0160296912206616)(-0.15,0.0160944604536453)(-0.14,0.0161520617583195)(-0.13,0.0161799886485759)(-0.12,0.0162303439761839)(-0.11,0.0162419483897245)(-0.1,0.0162887831305659)(-0.09,0.0163249538965256)(-0.08,0.0163779625833218)(-0.07,0.0163743888861845)(-0.06,0.0164110763847649)(-0.05,0.0164281296616598)(-0.04,0.0164179190993012)(-0.03,0.0164534504966913)(-0.02,0.016447467739857)(-0.01,0.0164632756502928)(0,0.0164565314948846)(0.01,0.0164695371948387)(0.02,0.0164500041263566)(0.03,0.0164622346427759)(0.04,0.0164242594296644)(0.05,0.0163993031613511)(0.06,0.0164127843520525)(0.07,0.0163640575896646)(0.08,0.0163654611795347)(0.09,0.0163338491633142)(0.1,0.016281838843329)(0.11,0.0162667313565729)(0.12,0.016213476626641)(0.13,0.0161726104207484)(0.14,0.0161165279574247)(0.15,0.0160866866017625)(0.16,0.0160235282545028)(0.17,0.0160063843247778)(0.18,0.0159286841273035)(0.19,0.0158802297841809)(0.2,0.0158189376169879)(0.21,0.0157427925993309)(0.22,0.015684557780372)(0.23,0.0155998759441254)(0.24,0.0155507660553255)(0.25,0.0155081075071791)(0.26,0.0154312914496241)(0.27,0.0153641631066202)(0.28,0.0152679195399679)(0.29,0.0152208716615959)(0.3,0.0151281327990724) 
};
\addlegendentry{$ \text{Sim. -- MLE} $};
\end{axis}
\end{tikzpicture}
\caption{$ \text{CRB} $ and simulated MLE MSE for GGD noise. Both the bound and simulated MSE were evaluated for a number of samples $ N=50 $ and for $ \varepsilon $ in the interval $ \left[-1,\;1 \right] $. The MSE for the MLE was evaluated using $ 10^{5} $ realizations of the sample blocks. We considered the following noise parameters: $ \beta=4 $ and $ \delta=1 $. }
\label{fig_ggd_small_n}
\end{figure}

To verify the concave behavior around $ \varepsilon=0 $ even in the cases where the CRB is not supposed to be tight, \textit{i.e.} for small $ N $, we simulated the MLE $ 10^{5} $ times for $ N=50 $ and the same GGD parameters previously used. The CRB and the simulation results are shown in Fig. \ref{fig_ggd_small_n}.

The results are given for a smaller interval $ \varepsilon\in\left[-0,3;0,3 \right]  $ when compared to the results for $ N=500 $  because, for a small $ N $, large $ \varepsilon $ may generate infinite estimates with a high probability ($ \vert \hat{X} \vert =+\infty $ when all the binary measurements have identical values). 

We can see that the MLE MSE still have the concave shape around $ \varepsilon=0 $, even if the bound is looser than for $ N=500 $.

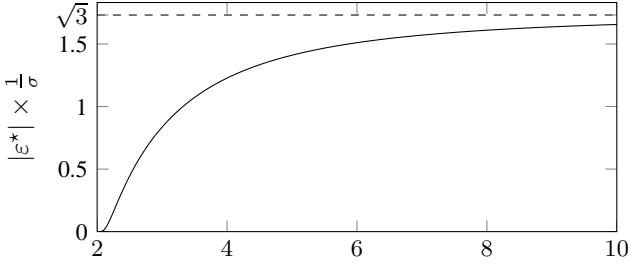
\begin{figure}[t!]
\centering
\begin{tikzpicture}
\begin{axis}[%
width=2.72in,
height=1.2in,
scale only axis,
xminorticks=true,
tick label style={font=\small},
ytick={0,0.5,1,1.5,1.73205080756888},
yticklabels={0,0.5,1,1.5,$ \sqrt{3} $},
label style={font=\small},
xlabel={$\beta$},
ylabel={$ \left\vert \varepsilon^{\star} \right\vert \times \frac{1}{\sigma}$},
ylabel near ticks,
xmin=2, xmax=10,
ymin=0, ymax=1.83205080756888,
axis on top]
\addplot [
color=black,
solid
]
coordinates{
 (2,0)(2.05,0.0001)(2.1,0.0098)(2.15,0.0451)(2.2,0.0978)(2.25,0.1576)(2.3,0.2188)(2.35,0.2789)(2.4,0.3367)(2.45,0.3918)(2.5,0.444)(2.55,0.4935)(2.6,0.5403)(2.65,0.5846)(2.7,0.6264)(2.75,0.6661)(2.8,0.7037)(2.85,0.7393)(2.9,0.7731)(2.95,0.8052)(3,0.8357)(3.05,0.8648)(3.1,0.8924)(3.15,0.9188)(3.2,0.9439)(3.25,0.9679)(3.3,0.9909)(3.35,1.0128)(3.4,1.0337)(3.45,1.0538)(3.5,1.073)(3.55,1.0915)(3.6,1.1092)(3.65,1.1261)(3.7,1.1424)(3.75,1.158)(3.8,1.1731)(3.85,1.1875)(3.9,1.2015)(3.95,1.2149)(4,1.2278)(4.05,1.2402)(4.1,1.2522)(4.15,1.2638)(4.2,1.2749)(4.25,1.2857)(4.3,1.2961)(4.35,1.3062)(4.4,1.3159)(4.45,1.3253)(4.5,1.3343)(4.55,1.3431)(4.6,1.3516)(4.65,1.3599)(4.7,1.3679)(4.75,1.3756)(4.8,1.3831)(4.85,1.3903)(4.9,1.3974)(4.95,1.4042)(5,1.4109)(5.05,1.4173)(5.1,1.4235)(5.15,1.4296)(5.2,1.4355)(5.25,1.4413)(5.3,1.4468)(5.35,1.4522)(5.4,1.4575)(5.45,1.4626)(5.5,1.4676)(5.55,1.4725)(5.6,1.4772)(5.65,1.4818)(5.7,1.4863)(5.75,1.4907)(5.8,1.4949)(5.85,1.4991)(5.9,1.5031)(5.95,1.507)(6,1.5109)(6.05,1.5146)(6.1,1.5183)(6.15,1.5219)(6.2,1.5253)(6.25,1.5287)(6.3,1.5321)(6.35,1.5353)(6.4,1.5385)(6.45,1.5416)(6.5,1.5446)(6.55,1.5475)(6.6,1.5504)(6.65,1.5533)(6.7,1.556)(6.75,1.5587)(6.8,1.5613)(6.85,1.5639)(6.9,1.5665)(6.95,1.5689)(7,1.5713)(7.05,1.5737)(7.1,1.576)(7.15,1.5783)(7.2,1.5805)(7.25,1.5827)(7.3,1.5848)(7.35,1.5869)(7.4,1.589)(7.45,1.591)(7.5,1.5929)(7.55,1.5949)(7.6,1.5967)(7.65,1.5986)(7.7,1.6004)(7.75,1.6022)(7.8,1.6039)(7.85,1.6056)(7.9,1.6073)(7.95,1.609)(8,1.6106)(8.05,1.6122)(8.1,1.6137)(8.15,1.6152)(8.2,1.6167)(8.25,1.6182)(8.3,1.6197)(8.35,1.6211)(8.4,1.6225)(8.45,1.6238)(8.5,1.6252)(8.55,1.6265)(8.6,1.6278)(8.65,1.6291)(8.7,1.6303)(8.75,1.6316)(8.8,1.6328)(8.85,1.634)(8.9,1.6351)(8.95,1.6363)(9,1.6374)(9.05,1.6385)(9.1,1.6396)(9.15,1.6407)(9.2,1.6418)(9.25,1.6428)(9.3,1.6438)(9.35,1.6448)(9.4,1.6458)(9.45,1.6468)(9.5,1.6478)(9.55,1.6487)(9.6,1.6496)(9.65,1.6505)(9.7,1.6514)(9.75,1.6523)(9.8,1.6532)(9.85,1.6541)(9.9,1.6549)(9.95,1.6557)(10,1.6566) 
};

\addplot [
color=black,
dashed
]
coordinates{
 (2,1.73205080756888)(10,1.73205080756888) 
};

\end{axis}
\end{tikzpicture}
\vspace{-15pt}
\caption{$\left\vert \varepsilon^{\star} \right\vert=\left\vert\underset{\varepsilon}{\arg\min}\,B\left(\varepsilon\right)\right\vert$ as a function of the shape parameter $ \beta $ for GGD with fixed standard deviation $ \sigma $.}
\label{fig_eps_beta}
\end{figure}

\paragraph*{Remark} when the noise PDF is flat around zero, the behavior of the bound seems to be linked to the fact that, asymptotically, the boundaries of the flat zone are very informative (the MLE for the location parameter of a uniform distribution depends on the maximum and on the minimum of the measurements). However, in the GGD case, the result seems less intuitive, as the uniform behavior is not present. In this case, it seems that the points of the minima define an equivalent uniform region, if we look from a location parameter estimation point of view. It is interesting to note, as it is shown in Fig. \ref{fig_eps_beta}, that for a fixed GGD standard deviation $ \sigma_{v} $, when we increase $ \beta $, the PDF gets closer to the uniform distribution and the two optimal $ \varepsilon $ tend smoothly to the boundaries of the uniform distribution $ \pm\sigma\sqrt{3} $.

\section{Effect of a noisy channel}
Until now, we considered that the estimator has direct access to the measurements. We can consider a more realistic situation where the sensor is distant and transmit the binary measurements through a noisy channel, for example a binary symmetric channel (BSC). The BSC changes the sign of a binary measurement with probability $ q<\frac{1}{2} $. Therefore, the estimator receives $ -1 $ with probability
\begin{equation}
r=q+\left(1-2q\right)F\left(\tau_{0}-x\right).
\label{eq_r}
\end{equation}
\noindent Similarly to the perfect channel case, we can use the frequency $ \hat{r} $ of receiving $ -1 $ to replace $ r $ in the expression above and then invert the function to obtain $ \hat{X} $:
\begin{equation}
\hat{X}=\tau_{0}-F^{-1}\left[ \frac{\left( \hat{r}-q\right)^{+}}{1-2q}\right], 
\label{eq_est_bsc}
\end{equation}
\noindent where the function $ \left(x\right)^{+}=\max\left(0,x\right)  $ is used because, in practice and specially for small $ N $, $ \hat{r} $ can be less than $ q $ (even if in theory it cannot) and $ F^{-1} $ is only defined for inputs in the interval $ \left[0,\,1\right]  $.

Also using a similar development as in Sec. \ref{est_var}, we obtain the following asymptotic variance or CRB
\begin{align}
\var\left(\hat{X}\right)\underset{N\rightarrow +\infty}{\sim} \text{CRB}\left(\varepsilon\right) =N^{-1 } B'\left(\varepsilon\right)\nonumber\\ 
\mbox{ with }
B'\left(\varepsilon\right)=B\left(\varepsilon\right)+\frac{q\left(1-q\right) }{\left(1-2q\right)^{2} }\frac{1}{f^{2}\left(\varepsilon\right) }.
\label{eq_b_prime}
\end{align}
Notice that for unimodal distributions for which optimal quantization is symmetric in the noiseless channel case, the introduction of a BSC does not make the optimal quantizer become asymmetric, as the additional term in $ B'\left( \varepsilon\right)  $ increases with $ \left\vert \varepsilon \right\vert $. However, for some unimodal distributions for which quantization is asymmetric in the perfect channel case, the introduction of a BSC can make optimal quantization become symmetric. We can easily see this effect through the condition for $ B'\left(0\right)  $ to be a local minimum:
\begin{equation}
-\frac{1}{\left(1-2q \right)^{2} }f^{\left(2\right)}\left(0\right)>4f^{3}\left(0 \right),   
\label{cond_bsc}
\end{equation} 
\noindent as $ \frac{1}{\left(1-2q \right)^{2} } $ is a strictly positive and increasing in $ q $, it can happen for some noise distributions that when $ q $ is zero or small the condition is not verified and we have asymmetric optimal quantization, then if the channel degrades, $ \varepsilon=0 $ becomes a local minimum and symmetric quantization becomes optimal. 

\section{Conclusions}
Differently from what is intuitively expected, we have shown in this letter that, when we estimate a location parameter based on binary quantized noisy measurements, the symmetry and unimodality of the noise PDF does not imply that the optimal quantizer must be symmetric, \textit{i.e.} with the threshold chosen to have equiprobable outputs. Even if this is true for commonly used noise PDF (Gaussian, Cauchy and Laplacian), we can find cases (uniform/Gaussian and GGD) where this is not only suboptimal but it is also locally the worst choice. This means that the optimal quantization strategy depends on the noise distribution and not only on the parameter being measured. Moreover, we verified that if we introduce a noisy channel (BSC) in the model, then the class of unimodal noise distribution for which optimal quantization is symmetric becomes larger as the channel noise is increased.


Finally, notice that asymmetry might also appear in detection of weak signals based on binary quantized measurements. In this case, it can be shown \cite{Kassam1977} that detection performance is directly related to the function $ B\left(\varepsilon\right)  $ studied here.



\section*{Acknowledgment}
The authors would like to thank Steeve Zozor for his helpful comments.


\ifCLASSOPTIONcaptionsoff
  \newpage
\fi

\bibliographystyle{IEEEtran}
\bibliography{biblio}

\end{document}